\documentclass[review]{elsarticle}

\usepackage{lipsum}
\usepackage{hyperref}
\usepackage[centertags]{amsmath}
\usepackage{amsfonts}
\usepackage{amssymb}
\usepackage{amsthm}
\usepackage{graphicx}
\usepackage{epsfig}
\usepackage{epstopdf}
\usepackage{float}
\usepackage{filecontents}
\usepackage{xcolor}
\usepackage[utf8x]{inputenc}
\input{amssym.def}

\theoremstyle{definition}

\usepackage{numcompress}

\begin{document}

\begin{frontmatter}

\title{A potential missile guidance law based-on chaos}

\author[]{Dhrubajyoti Mandal\corref{mycorrespondingauthor}}

\cortext[mycorrespondingauthor]{Corresponding author}

\ead{dmandal93@gmail.com}



\address{Department of Science $\&$ Technology

New Delhi-110016, India}

\begin{abstract}
An important field of research in defense-related technology is the design of guidance laws. A guided missile is generally challenging to intercept if its trajectory becomes unpredictable. In this short communication, we have discussed a possible application of the chaos theory in developing an advanced guided missile, where the guidance law is based upon a robust chaotic map. This type of guided missile may be almost impossible to intercept by existing missile defense systems due to its unpredictable trajectory. 
\end{abstract}
\begin{keyword} 
\texttt{Guided Missile; Chaos Theory.}
\end{keyword}

\end{frontmatter} 


\section{Introduction}

Chaos theory has found many applications in Physics \& Engineering \cite{awrejcewicz2012bifurcation, skinner1992application}. However, many more are yet to be explored. Chaos can occur in nonlinear continuous as well as discrete systems. One of the major characteristics of the trajectories of a chaotic system is that they remain bounded in a certain area of the phase space, although any two nearby trajectories separate from each other exponentially fast. Therefore, it is very difficult to predict the future dynamics of any trajectory of a chaotic system, although the system is deterministic in nature and the mathematical form governing the dynamics is already known.

Next, let us briefly discuss about guided missiles. Guided missiles are those missiles whose trajectory is determined by a certain fixed set of rules. In other words, guided missiles are given input at various instances during their flight to change their direction, velocity etc. This input is supplied according to some law, which is often known as 'Guidance Law'. Due to this guidance law, guided missiles do not follow simple projectile motion, rather, they follow a relatively complicated path to reach the target. Again, the external input determining the guidance law in a guided missile can be supplied in two different ways mainly. An input can be given from an external control point. Otherwise, an input can be supplied from some internal input generator or feedback system already installed in the missile, which is often called 'Inertial Navigation'. Extensive research on the guidance law of missiles has already been done in the literature \cite{li2014missile, manchester2006circular, pastrick1981guidance, turetsky2003missile, yanushevsky2018modern}. Missile guidance law based-on chaotic perception has also been proposed \cite{yu2009autonomous}. Also, there has already been extensive research in the literature on the defense system against guided missiles. However, If the trajectory of an incoming missile is highly unpredictable, then it is very difficult to determine the guidance law for the interceptor deployed for defense. Due to the aforesaid reason, guided missiles are a lot more difficult to intercept as they do not follow a pre-determined projectile motion.

\section{Main Idea}
In this short communication, we introduce the concept of a type of guided missile where the guidance law is based on chaos theory. It is well known that chaos occurs only in nonlinear systems. The motivation of our present work is to build an advanced guided missile where the direction and velocity of the missile will change at several points of its flight, and these changes will be determined by the dynamics of a robust chaotic map (Robust chaotic maps are those that exhibits chaotic dynamics for a wide range of one or more parameter values). As the trajectory of a chaotic map is highly unpredictable over time, therefore the resulting guided missile will follow an unpredictable path and this will make it difficult to intercept. In order to serve our purpose, we shall use a robust chaotic 2D map. There are many  robust two-dimensional chaotic maps in the literature. We can choose any such map to serve our purpose. Let us illustrate the above idea in more detail in the next section.

\section{Guidance law based-on chaos}

Suppose we want to make a guided missile which is carrying a nuclear weapon, and our aim is to hit a target located at a point, say 'A'. Now, as the missile is carrying a nuclear weapon we can allow a small error in hitting the target. That there exists a neighbourhood of this point, say, 'N(A)' so that even if the missile hits anywhere inside N(A), the goal is achieved. In that case, we have to make sure only that missile hits inside N(A) without any interception, and for that an almost unpredictable and complicated trajectory of the missile is essential.

To ensure above mentioned process, first we consider a chaotic map $f: \mathcal{B} \to \mathcal{B}$, where $\mathcal{B}$ is a compact subset of $\mathcal{R}^2$. Let, $g:  \mathcal{R}^2 \to \mathcal{R}^2$ be a continuous function that maps the compact subset $\mathcal{B}$ into $N(A)$ i.e. $g(B) \subset N(A)$. Then we consider an orbit of the map $\mathcal{O}_f(x)$ where $x=g(A)$. Obviously,  $\mathcal{O}_f(x) \subset \mathcal{B}$. Now we look at the corresponding image of the orbit under the transformation $g$, in other words we look at the set $g(\mathcal{O}_f(x))$ and clearly, $$g\left(\mathcal{O}_f(x)\right) \subset N(A).$$

Next suppose we wish to change the direction and velocity of the missile at 'n' different points in its entire path. For that, we choose a subset $T_n \subset g\left(\mathcal{O}_f(x)\right)$. This subset contains exactly '$n$' number of points inside $N(A)$. Let us denote, $$T_n =\{x_1, x_2, x_3, \cdots, x_n\}$$ where $$x_i \in g\left(\mathcal{O}_f(x)\right)$$ for $i=1,2,\cdots,n$. We call the points of $T_n$ as 'Variable Target Points'.

Next we discuss a guidance law based on the above formulation. Let a missile be launched in such a way that it will reach point $A$ via a projectile motion, if its motion remains unaltered for the total flight. Let the missile continue to go on and after a certain time it reaches to a point $P_0$ in the air. Then it changes the direction of motion and velocity so that  it could reach the point $x_1$ (this point is inside $N(A)$) via a projectile motion starting from $P_0$. Then, after sometime, when it reaches another point $P_1$ during its flight, it changes its direction and velocity in order to reach $x_2$ (this point is again inside $N(A)$), via a projectile motion starting from $P_1$. The same process is repeated and it continues to change its direction and velocity accordingly at $n$ different points during the whole flight. If we continue in this way, it is expected that the missile will ultimately change its direction and velocity to hit the point $x_n$ and it will explode after hitting the target point $x_n$. Here it may be noted that there may be restrictions regarding the change in velocity and direction of a missile at different points during the flight. However, we can always impose the condition that such changes will occur only if it remains within the permissible limit. The most crucial point to note in the above discussion is that the variable target points are determined by an orbit of a robust chaotic map. Due to this reason, the flight path of the missile becomes readily dependent on the chaotic dynamics of the chosen map.

\section{Advantages of the proposed guidance law}

We address he most important question here: What are the advantages of implementing the guidance law discussed above? The main advantage of this method is that the input for inertial navigation at every turning point need not be pre-determined. We only have to pre-define the chaotic map $f$, which exhibits robust chaos in a range of its parameter values. A chip located inside the missile can do the rest of the  mathematical calculations required. As the map $f$ exhibits robust chaos, i.e., exhibits chaos for a wide range of  parameter values. Now the whole flight path of the missile will depend on the particular value of the parameters chosen and on the particular finite subset $T_n$ of its orbit $\mathcal{O}_f(x)$.

Also it is important to note that we need not pre-define the parameter values or send any input about choices of particular parameter values from any control station. Even if the chosen subset $T_n$ of the orbit is not pre-determined. The parameter values will be automatically chosen by the system installed in the missile according to the environmental temperature sensing or atmospheric pressure sensing at a particular time during its flight, before reaching the first turning point on its path. We only have to pre-define a one-to-one function that maps an environmental temperature range or atmospheric pressure range to the range of the parameter values in which $f$ exhibits robust chaos and insert this information in the missile system before launch. This adds an extra advantage that before launching the missile, even the launcher will have no information about its flight path or trajectory. So there is no chance of information leakage about the possible trajectory of the missile. However, it is guaranteed that it will reach and hit the desired neighbourhood of the exact target and that will be enough. 

A disadvantage of the proposed guidance law is that it fails if it is  desired to hit a pre-determined exact target point. Moreover, notice that we must consider a robust chaotic map to ensure the existence of a parameter range over which the chaotic nature is not destroyed.

\section{Conclusion}

The theoretical approach discussed in this letter may be difficult to implement in practical models. We believe that extensive research is needed for further development of the proposed guided missile. Therefore, the work presented above is only a potential application of chaos theory in developing advanced guided missiles.

\section*{Funding} The author declares that no funding has been received for this research.
\section*{Conflict of Interest} The author declares that there is no financial or non-financial conflicts of interest.

\bibliography{missile}
\bibliographystyle{abbrv}
\end{document}